\documentclass[german,english]{bvm2019}

\usepackage{xcolor}
\usepackage[export]{adjustbox}
\definecolor{FAUblue1}{RGB}{0,56,101}
\definecolor{FAUblue2}{RGB}{144,167,198}
\definecolor{FAUblue3}{RGB}{221,229,240}

\definecolor{FAUgray1}{RGB}{152,164,174}
\definecolor{FAUgray2}{RGB}{210,213,215}
\definecolor{FAUgray3}{RGB}{235,236,238}

\definecolor{ltwonormcolor}{HTML}{989a9b}

\newcommand{\eccLineWidth}{0.3mm}
\newcommand{\plotMotionLineWidth}{0.25mm}

\def\etal.{et\penalty50\ al.}
\newcommand{\thd}{\mbox{3-D}~}

\newcommand{\consValwospace}{CIV}
\newcommand{\consVal}{\consValwospace~}
\newcommand{\consVals}{\consValwospace{}s~}

\usepackage{tikz}

\usepackage{ifthen}

\newboolean{doubleblind} 
\setboolean{doubleblind}{false} 

\title{Maximum Likelihood Estimation of Head Motion using Epipolar Consistency}

\ifdoubleblind
\else
\author{Alexander~Preuhs$^1$ \and Nishant~Ravikumar$^1$   \and Michael~Manhart$^2$  \and Bernhard~Stimpel$^1$ \and Elisabeth~Hoppe$^1$ \and Christopher~Syben$^1$ \and Markus~Kowarschik$^2$ \and Andreas~Maier$^1$}
\authorrunning{Preuhs et al.}
\institute{%
	$^1$Pattern Recognition Lab, Friedrich-Alexander-Universit{\"a}t Erlangen-N{\"u}rnberg\\
	$^2$Siemens Healthcare GmbH, Forchheim, Germany}
\email{alexander.preuhs@fau.de}
\fi

\begin{document}




%
\selectlanguage{english}

\maketitle

\DeclareRobustCommand\Tyoffset {\tikz[baseline=-0.6ex]\draw[draw=FAUblue2,line width = \eccLineWidth] (0,0)--(0.5,0);}
\DeclareRobustCommand\TrueGeo  {\tikz[baseline=-0.6ex]\draw[draw=FAUblue1, ,line width = \eccLineWidth, densely dotted] (0,0)--(0.5,0);}
\DeclareRobustCommand\Tzoffset{\tikz[baseline=-0.6ex]\draw[draw=FAUgray3,line width = \eccLineWidth, densely dashed] (0,0)--(0.54,0);}
\DeclareRobustCommand\Rxoffset{\tikz[baseline=-0.6ex]\draw[draw=FAUgray2,line width = \eccLineWidth,densely dashdotted] (0,0)--(0.54,0);}

\DeclareRobustCommand\gt {\tikz[baseline=-0.6ex]\draw[draw=FAUblue2,line width = 0.4mm] (0,0)--(0.5,0);}
\DeclareRobustCommand\loglikeli  {\tikz[baseline=-0.6ex]\draw[draw=FAUblue1, ,line width = \plotMotionLineWidth, densely dotted] (0,0)--(0.5,0);}
\DeclareRobustCommand\lone {\tikz[baseline=-0.6ex]\draw[draw=ltwonormcolor,line width = \plotMotionLineWidth, densely dashed] (0,0)--(0.52,0);}
\DeclareRobustCommand\ltwo{\tikz[baseline=-0.6ex]\draw[draw=FAUgray1,line width = \plotMotionLineWidth, dashdotted] (0,0)--(0.54,0);}

\begin{abstract}

Open gantry C-arm systems that are placed within the interventional room enable \thd imaging and guidance for stroke therapy without patient transfer. This can profit in drastically reduced time-to-therapy, however, due to the interventional setting, the data acquisition is comparatively slow. Thus, involuntary patient motion needs to be estimated and compensated to achieve high image quality. Patient motion results in a misalignment of the geometry and the acquired image data. Consistency measures can be used to restore the correct mapping to compensate the motion. They describe constraints on an idealized imaging process which makes them also sensitive to beam hardening, scatter, truncation or overexposure. We propose a probabilistic approach based on the Student's t-distribution to model image artifacts that affect the consistency measure without sourcing from motion.

\end{abstract}
\section{Introduction}
Modern C-arm systems enable \thd imaging of the head in an interventional environment. This is of high relevance in neuroradiology, where a \thd reconstruction allows to distinguish an ischemic from hemorrhagic stroke. The patients benefit from reduced time-to-therapy \cite{psychogios2017one} but the open gantry system and the interventional setting constrain the acquisition speed compared to conventional Computed Tomography (CT). With prolonged scan time, involuntary movements of patients constitute a major challenge for high quality image reconstruction.

This gives rise to a strong need for motion compensation algorithms \cite{muller2014image}. In recent years, consistency conditions have been shown to be promising in this context \cite{preuhs2018double,Frysch2015}. Besides the compensation of motion, consistency measures are heavily used in cone-beam CT, as they provide a mathematical model constraining the imaging process. The most commonly applied consistency measure uses Grangeat's theorem to judge the pairwise consistency of two projections. 

The measure was successfully applied for the correction of a variety of acquisition artifacts. Beam hardening can be corrected by projection linearization using a polynomial model. The parameters for the model are found by optimizing for Grangeat's consistency \cite{abdurahman2018beam,wurfl2017epipolar}. Hoffmann \etal. used Grangeat's consistency to estimate parameters of an additive scatter model \cite{hoffmannempirical}. The consistency measure can also be used to estimate missing data due to truncation for field of view reconstruction \cite{Punzet2018} and possibly to the closely related problem of overexposure correction \cite{preuhs2015over}. The most widely use of Grangeat's consistency is the estimation of geometry information that is distorted either due to rigid patient motion or geometry jitter \cite{preuhs2018double,Frysch2015,Debbeler2013}. The reason for its wide applicability is the sensitivity of the consistency measure to a variety of image artifacts. Thus, when we use the consistency measure to compensate for motion, we also measure the inconsistency induced by other sources. In this work, we propose to use a statistical model to handle inconsistencies that are not originating from motion artifacts.

\section{\consVal Look-Up-Table}
\begin{figure}
	\centering
	\resizebox{1\linewidth}{!}{
		\begin{tikzpicture}
			\input{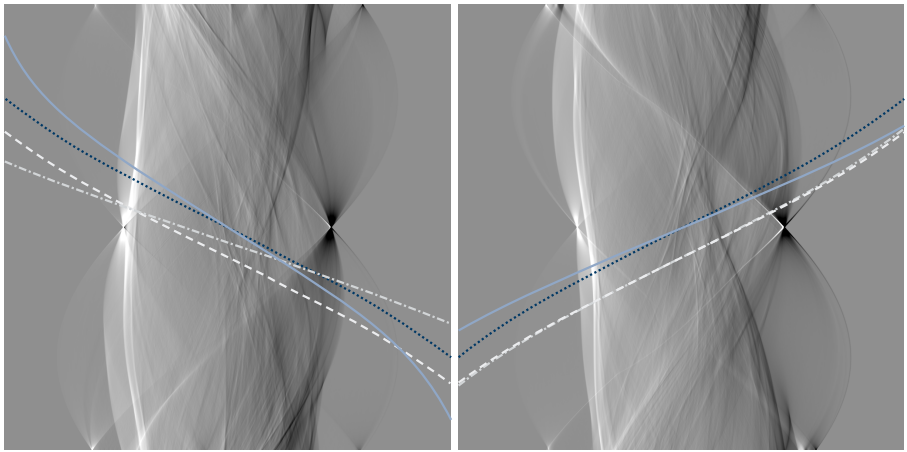}
		\end{tikzpicture}
	}
	\caption{LUT of two projections containing \consVals and corresponding lines visualizing the sampling of redundancies based on the given geometry information. The \consValwospace{}s along corresponding lines in the left and right LUT, e.g. along the dotted lines (\TrueGeo), should be equal (see Fig.\,\ref{fig:grangeatProfile} for a visualization of line profile). True geometry: dotted (\TrueGeo), $z$~translation: dashed (\Tzoffset), $x$~rotation: chain (\Rxoffset) and $y$~translation: solid (\Tyoffset).}
	\label{fig:derivativeRadon}
\end{figure}
In cone-beam imaging, a line profile $\vec{l}$ on the detector plane is created by attenuated X-ray beams within a plane $\vec{p}$ connecting the X-ray source and $\vec{l}$. Grangeat's theorem describes a transformation, to find a common value~---~we denote it as Consistency Intermediate Value (\consValwospace)~---~that can be computed either from $\vec{l}$, or the \thd Radon value indexed by $\vec{p}$ \cite{Aichert2016}. Formally, the derivative of the \thd Radon value in the normal direction of $\vec{p}$ equals a transformed value of $\vec{l}$. As a consequence, any pair of epipolar lines can be used as a consistency measure by computing their respective \consValwospace{}s, which must be equal. All  \consVals can be precomputed as a look-up-table (LUT) by a concatenation of cosine-weighting, Radon transform and derivative. As a result, from each projection we obtain a \consVal look-up-table as depicted in Fig.\,\ref{fig:derivativeRadon}. 

\begin{figure}[h]
	\centering
	\resizebox{\linewidth}{!}{
		\adjustbox{width=\linewidth,trim=10ex 0ex 0ex 0ex}{
			\input{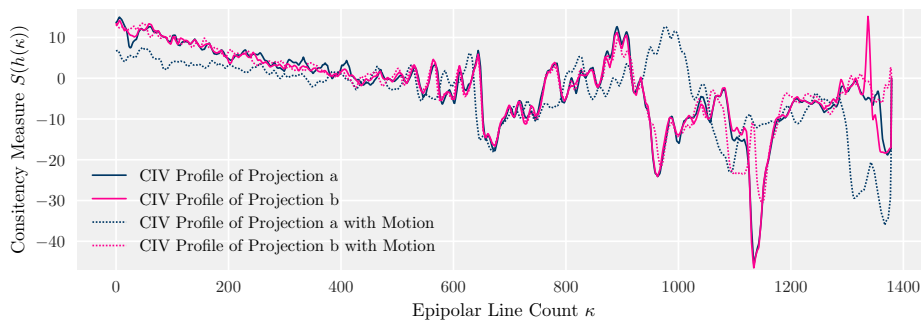}}}
	\caption{Profiles along the LUTs for correct and motion-affected geometry. The profiles are sampled along the true geometry and the geometry affected by a translation in $y$ direction as depicted in Fig.\,\ref{fig:derivativeRadon}.}
	\label{fig:grangeatProfile}
\end{figure}

Using the geometry information, we can sample \consVals \cite{Aichert2016} as visualized in Fig.\,\ref{fig:derivativeRadon}. Here, the dark blue dotted line corresponds to a correct sampling, where the geometry information is consistent with the acquired projection data. The profile of the sampled line is shown in Fig.\,\ref{fig:grangeatProfile}. If the geometry information is corrupted then the sampling pattern changes, as visualized by the solid line in Fig.\,\ref{fig:derivativeRadon}. The corresponding sampling profile is shown in Fig.\,\ref{fig:grangeatProfile}. Due to the misalignment, the profiles sampled from both LUTs, corresponding to two projections, do not match anymore. This is used to restore the correct geometry by minimizing the difference between two profiles and in turn maximizing the consistency. However, the profiles of two projections with perfect geometry will not match exactly (cf.\,Fig.\,\ref{fig:grangeatProfile}), because other acquisition artifacts reduce the consistency. We propose to model these artifacts using a probabilistic approach. 

\section{Student's t-Distribution-based Maximum Likelihood Estimation for Consistency Optimization}
\begin{figure}
	\centering
	\resizebox{0.9\linewidth}{!}{
		\adjustbox{width=\linewidth,trim=10ex 0ex 0ex 5ex}{
			\input{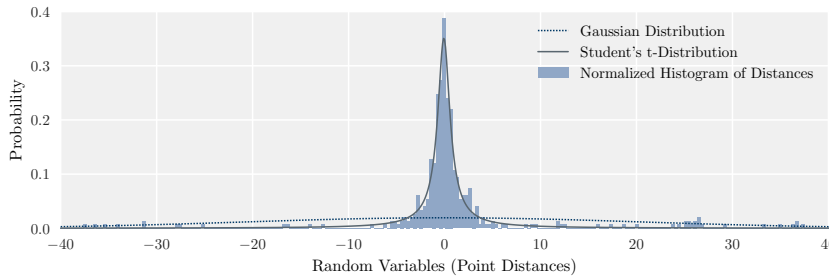}}}

	\caption{Histogram of distances between corresponding \consVals from two projections (cf.\,Fig.\,\ref{fig:grangeatProfile}) and fitted probability density functions.}
		\label{fig:fitting}
\end{figure}

Instead of assuming two \consVals to be equal, we propose a Bayesian description of the matching problem. We assume that the sampled \consVal of projection $a$ is a random variable $x$, distributed according to a Student's t-distribution with mean $\mu$, variance $\sigma$ and a shape factor $\nu$
\begin{equation}
p(x|\mu,\sigma,\nu) = \frac{\Gamma(\frac{\nu+1}{2})}{\sqrt{\nu\,\pi}\,\Gamma(\frac{\nu}{2})\,\sigma} \left(1+\frac{(x-\mu)^2}{\nu \,\sigma^2}\right)^{-\frac{\nu+1}{2}} \enspace ,
\end{equation}
where $\Gamma$ is the gamma function. We use a t-distribution, because it can model our inherently outlier-affected sampling process (cf.\,Fig.\,\ref{fig:fitting}). We assume that the mean of the random variable is the \consVal of our second projection image $b$ described by $S_b(h(\kappa))$. Here $S_b$ is the LUT generated from projection $b$, and $h(\cdot)$ defines a function that maps an angle $\kappa$ to a corresponding value in the LUTs as displayed in Fig.\,\ref{fig:derivativeRadon}. A detailed explanation on the function $h(\cdot)$ can be found in \cite{Aichert2016}. 

The motion compensation can then be formulated by finding the parameters of a probability density function that results in the greatest likelihood, or alternatively minimizing the negative log-likelihood. By setting the random variable $x$ as $S_a(h(\kappa))$ the maximum likelihood is defined by
\begin{equation}
\min\, \sum_\kappa -\ln\left(p(x|\mu,\sigma,\nu)\right) \,= \,\min\, \sum_\kappa \frac{\nu+1}{2} \, \ln\left(1+\frac{\left(S_a\left(h(\kappa)\right)-S_b\left(h(\kappa)\right)\right)^2}{\nu\,\sigma^2}\right) \enspace.
\label{eq:loglikelihood}
\end{equation}
For similar objects (e.g. head) and a given system, the distribution will always be similar. Thus, the unknown parameters $\nu$ and $\sigma$ can be estimated from a motion free prior scan (not identical to the scanned object) by accumulating the distances within the \consVals given by $\sum_{\kappa} S_a(h(\kappa))- S_b(h(\kappa))$. In a second step we fit a Student's t-distribution to the data. This is depicted in Fig.\,\ref{fig:fitting}, where we can see that the t-distribution properly fits to the data: outliers are modeled by its approximately constant tails. 
For comparison, fitting a Gaussian distribution leads to a very high standard deviation ($\sigma = 29$) due to the outliers.%
\section{Experiments and Results}
\begin{figure}
	\centering
	\resizebox{\linewidth}{!}{%
		\input{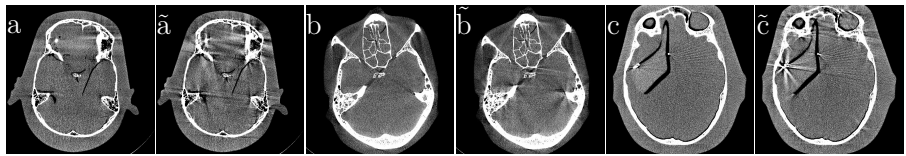}
	}%
\label{fig:recos}
\caption{Central slices of three reconstructed anthropomorphic head phantoms (HU \mbox{[-200,400]}). Motion-compensated reconstruction using the proposed method ($a$,$b$,$c$) and reconstruction with motion artifacts ranging from $\pm 0.5$ mm ($\tilde{\text{a}}$,$\tilde{\text{b}}$,$\tilde{\text{c}}$).}
\end{figure}

\begin{figure}[h]
	\centering
	\resizebox{0.95\linewidth}{!}{
		\adjustbox{width=\linewidth,trim=5ex 0ex 0ex 0ex}{
			\input{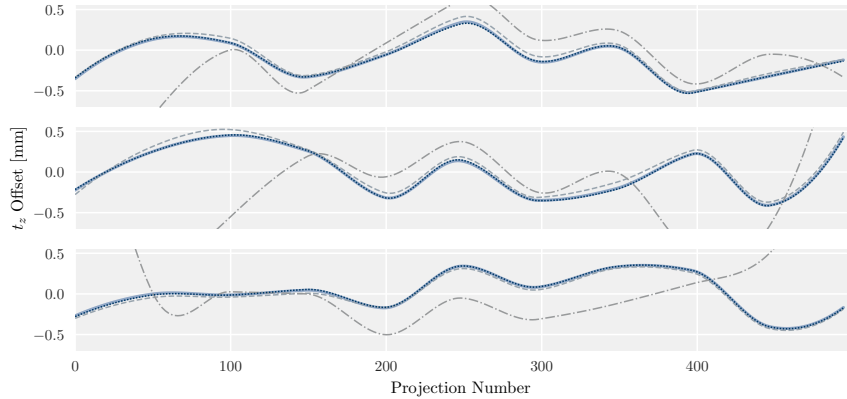}
		}
	}

\caption{Splines describing motion trajectory for datasets a (top), b (middle) and c (bottom). Each subplot shows the ground truth motion trajectory (\gt) and the estimated trajectory using the proposed norm (\loglikeli), the $L^1$ norm (\lone) and the $L^2$ norm (\ltwo).}
\label{fig:motionCurves}
\end{figure}

~\\
\textit{Experimental Setup:}
To evaluate our method, we compare our norm, derived as the log-likelihood of a Student's t-distribution ($\sigma=0.398$, $\nu=0.8228$) with the $L^2$ norm, which is the log-likelihood assuming a Gaussian distribution and the more robust $L^1$ norm. We apply them for the compensation of axial motion ($z$) modeled with splines. Thus, each projection is shifted in the $z$ direction by the amount of the spline at that projection index. The spline shape is controlled by 12 nodes, whose values are determined randomly in the range of $\pm 0.5$ mm. We use $z$-motion, because $z$-translations produce inconsistencies. In contrast, motions within the acquisition plane do not necessarily violate the consistency measure \cite{preuhs2018double,Frysch2015}. We show our results on three phantoms (cf.\,Fig.\,\ref{fig:recos}), acquired with a robotic C-arm system (Artis zeego, Siemens  Healthcare GmbH, Germany).
~\\
\textit{Motion Compensation:}
We use a simplex method to find the motion-compensated geometry. We iteratively optimize each node of a spline separately, assuming all other spline nodes constant. For each dataset, we model a different motion trajectory. The induced motion trajectory is displayed in Fig.\,\ref{fig:motionCurves} for all the three datasets, with the estimated motion curves using the respective norms. 
~\\
\textit{Results:}
The motion-estimated reconstruction using the proposed method is displayed in Fig.\,\ref{fig:recos} together with the motion-affected reconstruction. The mean distance between the estimated and ground truth motion parameters are displayed in Tab.\,\ref{tab:motionMean}. Using the proposed norm, we achieve the best results in restoring the motion parameters. The second best estimation is achieved using the robust $L^1$ norm. Using the $L^2$ norm the consistency and the motion parameters can only be poorly approximated. A visual inspection of the parameters is provided in Fig.\,\ref{fig:motionCurves}. The $L^2$ norm especially fails to approximate the areas at the beginning and end of the trajectory. The motion structure in the middle is approximated in its structure, although, with a great offset.

\begin{table}
	\centering
	\caption{Mean distance [mm] between estimated and ground truth motion curves.}
		{\def\arraystretch{1}\tabcolsep=13pt
	\begin{tabular}{c|ccc}
		\hline 
		&  			Dataset a&  Dataset b	& Dataset c  \\ 
		\hline 	\hline
		Proposed&  \textbf{	0.00567}		&  	\textbf{	0.00444}	& \textbf{0.00477}\\ 
		\hline 
		$L^1$&  	0.01521		&  			0.01579		& 0.00772 \\ 
		\hline 
		$L^2$&  		0.50440	&  		0.61670		&  0.42677 \\ 
		\hline 
	\end{tabular}}
	\label{tab:motionMean} 
\end{table} 
\section{Conclusion}
We propose a statistical description for evaluating the consistency of a trajectory. We model the consistency value as a Student's t-distribution and find the optimal geometry by minimizing the negative log-likelihood. Consequently, we derive a robust norm for the comparison of consistency values, insensitive to outliers which naturally arise due to physical effects. The proposed solution outperforms the $L^1$ and $L^2$ norm. The $L^2$ norm is very sensitive to outliers which pose it improper. In the current approach, we fixed the parameters of the t-distribution. Each projection pair reveals a different outlier-characteristic due to the scanned object. Thus, the precision might be enhanced by estimating projection pair dependent parameters for the Student's t-distribution.  
~\\
\textbf{Disclaimer:} The concepts and information presented in this paper are based on research and are not commercially available.
\bibliographystyle{bvm2019}
\bibliography{0000}
\end{document}